\documentclass[twocolumn,showpacs,preprintnumbers,amsmath,amssymb]{revtex4}
\usepackage{graphicx}
\usepackage{amsmath}
\usepackage{amsfonts}
\newcommand{\be}{\begin{equation}}
\newcommand{\en}{\end{equation}}
\newcommand{\no}{\nonumber}
\newcommand{\fr}[2]{\frac{#1}{#2}}
\newcommand{\q}{\quad}
\begin{document}
\title{Geometry of density states}
\author{Luis J. Boya}
\email {luisjo@unizar.es}
\thanks{Permanent address: Departamento de F$\acute{\textrm{i}}$sica Te$\acute{\textrm{o}}$rica. Universidad de Zaragoza.
E-50009 ZARAGOZA, Spain}
\author{Kuldeep Dixit}
\email{kuldeep@physics.utexas.edu}
 \affiliation{Department of Physics,
        University of Texas,
        Austin, TX 78712.}

\begin{abstract}
We reconsider the geometry of pure and mixed states in a finite
quantum system. The ranges of eigenvalues of the density matrices
delimit a regular symplex  (Hypertetrahedron $T_N$) in any
dimension $N$; the polytope isometry group is the symmetric group
$S_{N+1}$, and splits $T_N$ in chambers, the orbits of the states
under the  projective group $PU(N+1)$. The type of states
correlates with the vertices, edges, faces, etc. of the polytope,
with the vertices making up a base of orthogonal pure states. The
entropy function as a measure of the purity of these states is
also easily calculable; we draw and consider some isentropic
surfaces. The Casimir invariants acquire then also a more
transparent interpretation.
\end{abstract}

\pacs{03.65.Ca, 02.10.De, 02.40.Et}

\date{\today} \maketitle

\section{Introduction}
\textit{Pure} states  $\pi$ in Quantum Mechanics are described by
rays in a Hilbert space $\mathcal{H}$, or equivalently as
unidimensional subspaces. As such, they can be identified with
orthogonal projectors of unit trace (so for $k$-dimensional
subspaces they will have trace $k$); denoting by End $V$  the
linear operators of a vector space $V$, we have
\begin{multline}
\{\mbox{space of pure states}\} \approx \{\pi \in \mbox{End }
\mathcal{H} |\; \pi =\pi^\dagger, \\
\quad \mbox{Tr} [\pi] = 1,
\quad \pi^2 = \pi\}
\end{multline}

To include mixed states $\rho$, it is enough (von Neumann) to
relax idempotency  ( $\pi^2 = \pi$, so Spectrum $\pi  = \{1, 0\}$)
to mere positivity, $\rho \geq 0 \leftrightarrow$ Spec $\rho \geq$
0, which implies   $0 \leq \mu_i \leq 1$ for $\mu_i \in$ Spec
$\rho$. Hence
\begin{multline}
 \{\mbox{space of pure/mixed states}\} \approx
\{\rho \in \mbox{End } \mathcal{H} |\; \rho =\rho^\dagger, \\
\quad \mbox{Tr} [\rho] = 1, \quad \rho \geq 0\} \end{multline}

 So the pure states are \textit{extremal}, as $\rho^2 \leq
\rho$, and one can write the mixed states as incoherent addition
of pure ones,

\begin{equation}
\rho=\sum p_i \pi_i, \quad \sum p_i=1, \quad  p_i \geq 0
\end{equation}

For finite quantum systems ( $\mathcal{H} \rightarrow V$, dim $V =
N < + \infty$ ) the geometry of these states has been studied
lately from different points of view, as it is pertinent to
quantum information and other modern applications. For example, in
\cite{1} the nature of the space of pure states, namely the
projective spaces $CP^{N-1}$, is stressed, as they are rank-one
symmetric spaces; in \cite{2} the mixed states are related to a
coherence vector $\overrightarrow{n}$, lying in a subset of
euclidean space, determined by the values  of Casimir invariants
for the group $SU(N)$. \cite{3} studies the stratification of
general states under the projective group

\begin{equation}
         PU(N) = U(N)/U(1) = SU(N)/Z_n
\end{equation}
in relation to a (in general not regular) foliation by
K$\ddot{\textrm{a}}$hler manifolds, similar to the orbits of the
(co-)adjoint representation of $SU(N)$, that is, the
Kirillov-Souriau method. The geometry of the $N=3$ case had been
studied sometime ago also by Michel \cite{4} in relation to
$SU(3)$ as a flavor group.

In this report we would like to complete the above descriptions in
several ways. In particular, we elaborate an idea in \cite{3},
showing that the set of eigenvalues  $\mu_i$  of the density
operator $\rho$, $\mu_i \in$ Spec $\rho$, coincides precisely with
the points of a solid simplex (the hypertetrahedron $T_N$), the
simplest regular polytope. The natural isometry group, the full
symmetric group $S_{N+1}$, corresponds, of course, to a (finite)
Coxeter group, that is, the Weyl group of the Lie algebra $A_N$ of
the group $SU(N+1)$: the unit trace in our case is traded for the
traceless character of (anti-)hermitian matrices describing $A_N$.
The orbit space $T_N/S_{N+1}$ is still an (irregular)
hypertetrahedron, of size $1/(N+1)!$ of the previous regular one,
behaves like the Weyl chambers, and describes precisely the orbits
of our states under the projective group, each point being just an
orbit. The type of orbits, as classified by the little groups, is
in correspondence with the combinatorial elements of the
simplices, namely vertices, edges, faces and so on, and also with
the partitions of the number $N$.

For example for $N=2$ the set of pure states is the 2-sphere
$S^2=CP^1$, the set of all states is the three dimensional Bloch
ball B$^3$ of unit radius, with boundary being the pure states,
the simplex is just the closed segment  I: [-1, 1], the Weyl group
Z$_2$ is reflection in the middle, and the chamber is the
half-segment I/Z$_2$ =[0, 1] with 0 the most mixed states (called
$mixMax$ or $mM$ henceforth), and with 1 indicating the (sphere
of) pure states. In this simplest (and nonrepresentative) case
there is only one stratum, i.e. one type of orbit (namely
2-spheres), besides of course the fix point $mM$ or 0; the
(single) Casimir $I_2$ just labels the radius of these spheres
$S_r$ , $0 < r\leq 1$. The $mM$ state corresponds to partition
[$2$], the rest to [$1^2$], as stratum orbits $\approx
U(2)/U(1)^2$.

The Casimir invariants $I_i$ admit a double interpretation, as an
homogeneous system of generators of the center of the enveloping
algebra of Lie group, or in our case, as the symmetric functions
(Newton) over the roots of the spectral equation for the density
operator. We elaborate here the considerations of \cite{2}, based
on the pioneer work of Biedenharn \cite{5}.

    In this picture it is also clear how to compute the entropy $\eta$ of these mixed states, where
    $\eta(\rho)$ = - Tr$\rho$log($\rho$). This varies between $\eta$ = log $N$ for the $mM$ state to
    log $1$=0 for pure
    states. We include some graphs showing the entropy for some boundary lines on the "chambers" of the orbit
    space, as well as some isentropic surfaces in the general  case. Each case $N$ includes, in a precise sense,
    all the previous ones, $n < N$.

We stress also the action of the projective isometry group
$PU(N+1) \subset SO((N+1)^2-1)$ as the group acting effectively in
the space of states, in the sense of  the characterization of
geometries (F. Klein): in the transformation  $\rho \rightarrow
U\rho U^\dagger$ the effective group acting is $PU=U/U(1)$: the
kernel of the $U$ action is $U(1)$, and even $SU$ acts with kernel
$Z_n$: the explanative diagram is \be
\begin{array}{ccccc}
  Z_n & \rightarrow & SU(N) & \rightarrow & PSU(N) \\
  \downarrow &  & \downarrow &  & || \\
  U(1) & \rightarrow & U(N) & \rightarrow & PU(N) \\
  \updownarrow &  & \downarrow &  &  \\
  U(1) & == & U(1)  &  &  \\
\end{array}
\en The organization of the paper is as follows. In Sect. 2 we
establish notation and show several properties of our objects
needed later. In Sect. 3 we recall the situation for qubits
($N$=2), qutris ($N$=3), and $N$=4, incorporating the new
observations of above and suggesting generalizations to higher
dimensions. Then Sect. 4 deals with the general theory for
arbitrary $N$: we exhibit explicitely the polytope, the Weyl
quotient ("chamber"), the types of states, the Casimir invariants
and discuss the entropy function, and also some cases of
isentropic surfaces.

Other considerations (including mention of omissions) are in our
final Sect. 5.

\section{GENERAL DESCRIPTION AND PROPERTIES OF MIXED STATES}
In this Section we sumarize important results of \cite{2} to
\cite{4}). Let $\mathcal{S}$ be set of $N$-dimensional density
matrices $\rho$,
    and let $\mathcal{P}\subset  \mathcal{S}$  the subset of pure states,
    $\pi$, with
    $\pi^2=\pi$. The conditions

\begin{equation}
\rho=\rho^\dagger, \quad Tr [\rho] =1
\end{equation}
are respectively  linear and affine, so the whole set of solutions
of the equation is like a real vector space of dim $N^2$-1. The
positivity condition $\rho > 0$, though, is a $\textit{convex}$
condition, and selects a nonlinear submanifold $\mathcal{S}$ of
R$^{N^2-1}$ (which is closed as the spectral restriction is really
$0 \leq \mu_i \leq $ 1), of the same dimension, with $\mathcal{P}$
as the extremals of the convex set (not the boundary, in general:
we know of course that $\mathcal{P} = CP^{N-1}$, so dim
$\mathcal{P}  = 2N-2 \ll N^2-1$ for large $N$). In the set of
hermitian traceless matrices $\{h\}$ one introduces the definite
scalar product
\begin{equation}
            <h_1,h_2> := (1/2) \mbox{Tr} [h_1 h_2]
\end{equation}
and therefore there exists an orthonormal base
$\lambda_1$,$\lambda_2$, ..., $\lambda_{N^2-1}$ with properties
\begin{equation}
        \lambda_i = \lambda_i^\dagger,  \quad  \mbox{Tr}[ \lambda_i] =0,  \quad   \mbox{Tr} [\lambda_i^2] =
        2.
\end{equation}
So the general density matrix can be written as
\begin{equation}
\rho=\frac{1}{N}\left(I_n+\sqrt{\frac{N(N-1)}{2}}
\vec{n}\cdot\vec{\lambda}\right),
\end{equation}
where the factor  $\sqrt{N(N-1)/2}$ guarantees that pure states,
$\pi^2 = \pi$ have norm $|n|=1$; see \cite{2} for details;
$\vec{n}$ is called the $\textit{coherence}$ vector.

We recall also two related properties of the group $SU(N)$; first,
the square of the adjoint representation contains the adjoint BOTH
in the symmetric and in the antisymmetric part (for $N$=2 the
first is missing), and second, consequently, this induces two
algebras in the space of matrices (corresponding to the
coefficients $f_{ijk}$ and $d_{ijk}$ of Gell-Mann; of course, the
$f$'s correspond to the Lie algebra structure); for details
    see \cite{4}.

Now the group $PU(N)$ is acting in the set of states $\mathcal{S}$
as  $\rho \rightarrow U \rho U^{-1}$. If the eigenvalues of $\rho$
are different, the little group is $U(1)^{N-1}$, and then the
generic orbit is $PU(N)/U(1)^{N-1}$, of dimension $N$($N$-1). This
is a K$\ddot{\textrm{a}}$hler manifold, as are all orbits of the
adjoint of any simple Lie group (we notice that $N$($N$-1) is
even); in particular this space is called a (complex) flag
manifold, with structure (as homogeneous space),
\be
            \mathbb{F}l_C (N) = PU(N)/ U(1)^{N-1} = U(N)/U(1)^N.
\en

The other extreme contains the pure states, with little group
$U(N-1)$; these most critical orbits have dimension 2($N$-1), as
we said above. There is always naturally a single most-mixed
state, O (or $mM$), with $\vec{n}$ = 0: this is the fixed point
under the $PU$ action, and it is unique. Orbits in state space
with conjugate little groups form what is called a
\textit{stratum} (see e.g. \cite{4}), and in fact for a compact
group (as is our case) acting "nicely" in some space X the number
of strata is finite; we shall see that it  coincides with the
number of types of states and also naturally with the partitions
of $N$.

We define the entropy $\eta = \eta(\rho)$ of a state $\rho$ as the
expectation value of the operator $H$= -log ($\rho)> 0$ (von
Neumann; notice $\rho \leq 1$, so log($\rho$) is negative);
recalling that in the density formalism $<A> = \textrm{Tr} (\rho A
)$, we have, with $\mu_i \in$ Spec $\rho$,
\begin{equation}
           \eta (\rho) = - \textrm{Tr} \rho \log (\rho) = \log (\Pi
           \mu_i^{-\mu_i}).
\end{equation}
    This varies, as said, between log $N$ and log 1=0: the $mM$ state is the most disordered, and the pure states are the
    most
    ordered: we shall see also the $N$ case reproduces the entropy function for all previous cases $n <$ $N$, and
    therefore there must be several manifolds of mixed states \textit{isentropic}, with the same value for the entropy:
     we shall provide some examples.

    As for the Casimir invariants, we define them from the spectral equation for the density operator as the
    coefficients of the powers, i. e. :

\begin{equation}
    \rho^N - (\textrm{Tr} \rho)\rho^{N-1}+ I_2 \rho^{N-2} - I_3 \rho^{N-3}+ ... (-1)^N det \rho =0
\end{equation}
where $I_1$ = Tr $(\rho) \equiv 1$, (the first Casimir).  They
determine the spectrum up to a permutation, e.g.

    $I_2 = \sum_{i < j} \mu_i \mu_j$,        and    $I_n = \Pi \mu_i = \mu_1...\mu_N$

    These operators can be expressed also as traces of powers of representative matrices, as for example

\begin{eqnarray}
I_2 = (1/2)[(\textrm{Tr} \rho)^2 - \textrm{Tr}( \rho^2)]   ;
\mbox{ so it's zero iff }
\rho \mbox{ is pure}. \no \\
        I_3= (1/6)[(\textrm{Tr} \rho) ^3 +2 \textrm{Tr} (\rho^3) - 3 \textrm{Tr} (\rho) \textrm{Tr}
        (\rho^2)].   \nonumber
\end{eqnarray}
    There are also several inequalities assuring all Casimirs are nonnegative, etc.

\section{THE SIMPLEST CASES}

\begin{figure}
  \includegraphics[width=8 cm]{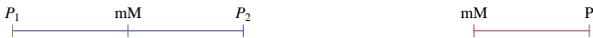}\\
  \caption{"(color online)" Eigenvalue segment for $N=2$ }\label{fig:2d}
\end{figure}

For $\textbf{N=1}$ we have just a single point, a pure state, as
$CP^0$ = {point}, with
        entropy log(1)= 0. For $\textbf{N=2}$ the density matrix can be written as
\be
                \rho =(1/2)(1 + \vec{\sigma} \cdot \mathbf{x});
\en
 we have  $\mathbf{x} \in$ $B^3$ ball (radius 1) with the 2-dim boundary sphere  $S^2=CP^1$ of pure states. The set of eigenvalues
 $\{(1+z)/2, (1-z)/2\}$ or  ($x, y$) with $x+y=1$ makes up a segment
 I in fig \ref{fig:2d},
where $P_1$ is e.g. the state [0, -1], $mM$ is (1/2, 1/2), and
$P_2$ is [1, 0]. The crucial but trivial point now is that under
the symmetry $Z_2$: ($x, y$) $\rightarrow$ ($y, x$)  the  two
states are equivalent, so the segment I becomes just the
half-segment  [0, 1]; that is, the $mM$ state (0), the mixed
states ($0 <x <1$) and the (representative of) pure state(s) $P$,
$x$= 1. In this case $PU(2)=SU(2)/Z_2=SO(3)$, so the orbits are
just spheres, and x in the half-segment means just the radius of
them. There are only two strata, the fixed point (little group
$SO(3)$) and the rest, little group SO(2)=U(1)). The two strata
correspond to the two partitions namely $U(2)/U(2)=\{Point\}=$$mM$
is [2] and $U(2)/U(1)^2=PU(2)/U(1)=CP^1$ is the partition
$[1^2]=[1,1]$.

\begin{figure}
  \includegraphics[width=6 cm]{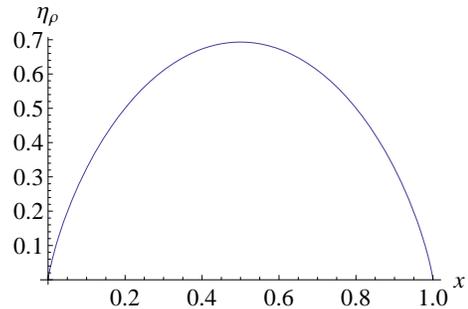}\\
  \caption{"(color online)" Entropy for $N$=2 case}\label{fig:2dentropy}
\end{figure}

The entropy is a smooth function from log(2)=0.693 (for the $mM$
state) to log(1)=0 for the pure one (fig \ref{fig:2dentropy}): we
shall see that the entropy function of the $N$ case always
contains that of the $n < N$ previous cases.

    There is only a Casimir, the quadratic one, with
\be
    I_2 =[(1+z)/2][(1-z)/2] =xy=x(1-x),
\en
which lies between 1/4 and zero.

    One can also use an "angle" picture \cite{2}, namely with $z=\cos\theta$, $0 \leq \theta \leq \pi$,
$x=\cos^2 \theta/2, y = \sin^2\theta/2$. The (quadratic) Casimir
is now $(1/4)\sin^2 \theta$ . The entropy in terms of angle
variables is
\begin{equation}
     \eta(\theta) = - \cos^2(\theta/2) \log[\cos^2(\theta/2)] - \sin^2(\theta/2) \log[\sin^2(\theta/2)
     ].
\end{equation}

\begin{figure}
  \includegraphics[width=6 cm]{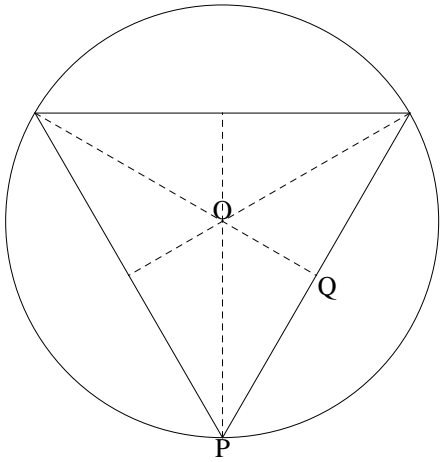}\\
  \caption{"(color online)" Positivity domain for $N$=3}\label{fig:3d}
\end{figure}

For $\textbf{N=3}$ we have $\rho = (1/3)(1 +  \sqrt{3} \: \bf{n}
\cdot \overrightarrow{\lambda}$ ), where the $\lambda$'s are e.g.
the Gell-Mann
    matrices. Choosing
    $\lambda_3$ and $\lambda_8$ diagonal, we get, with $n
    =(0,0,a,0,0,0,0,b)$ and
\begin{equation}
                                   \rho_{diag} = \fr{1}{3}\left(%
\begin{array}{ccc}
   1+\sqrt{3}a +b& 0 & 0 \\
  0 & 1-\sqrt{3}a +b & 0 \\
  0 & 0 &  1-2b\\
\end{array}%
\right)
\end{equation}
as $\rho_{diag} = (1+\sqrt{3}( a\lambda _3 +b\lambda _8))/3$.
Hence positivity implies $-1 \leq b \leq 1/2$, and one gets a
regular triangle "upside down" inscribed in the circle of radius
one (fig \ref{fig:3d}).

    Notice the three vertices $P_{1,2,3}$;  3 edges $l_{1,2,3}$, and the three interior lines (heights) $h_{1,2,3}$
    with intersection points $Q_{1,2,3}$. The figure is identical to the so-called Fano plane, related (among other
    things) to the octonion multiplication rule \cite{6} and to the projective plane $\mathbb{F}_2P^2$,with automorphisms
    the simple group $PSL_3(2) $of 168
elements \cite{7}. This is another test of the relation of $SU(3)$
with octonions!

    Under the action of the symmetry group, now $S_3$ with 6 elements, isomorphic to the dihedral group $D_3$ (the
    isometry of the regular n-gon is $D_n$), the fundamental domain or chamber is a little rectangular triangle
    (there are six of them), 1/6 in area of the big one; we select one of them
    as $\triangle OQP$ as in fig. \ref{fig:3d}. Here $O$ is the $mixMax$ state, $P$ the pure, $Q$ intermediate, i. e.,
                      $\eta(O)$=log(3), $\eta(Q)$=log(2), $\eta(P)$= log(1)=0.

   In terms of the eigenvalues of $\rho$, $O$ is (1/3, 1/3, 1/3), $P$ is (1, 0, 0), $Q$ is (1/2, 1/2, 0); recall that the order is immaterial.
The line $OQ$ is type $h$ ($xxy, \; x \geq$y); the line $QP$ is
type $l$  ($xy0, \; x \geq y \geq 0$), and this line reproduces
the $N$=2 case, with $Q$ seen as the $mM$ state.

The line $OP$ is type $h$, with ($xyy, \; x \geq y$). Note all the
statements are order-invariant! In fact in the large triangle the
line $POQ$ is a single line! Notice also the boundary lines in the
big triangle unite pure states, so they have 0 in the spectrum.
Each line represents the three embeddings of $SU(2)$ in $SU(3)$,
called by Gell-Mann $u,\; v$ and $w$ spin. It becomes a single
line in the little triangle, line $PQ$.

Note the set of generic states (little group $U(1)^2$) is
6-dimensional, open and dense; therefore, the representatives are
bi-dimensional (2+6=8=dim $SU(3)$).  The critical states (two
equal eigenvalues, little group $U(2))$ are the two lines
emanating from $O$, but $O$ is the fixed point, with isotropy just
$PU(3)$ itself. The three partitions of [3] are: $ mM= U(3)/U(3)$
is [3], $U(3)/U(1)\times U(2)=CP^2$ and $U(3)/U(1)^3$, which is
the flag manifold. Notice also the little triangle is rectangular,
NOT regular.

The Casimir invariants now are two, quadratic and cubic. They are
\be
       I_2=\mu_1\mu _2+ \mu_2\mu _3+  \mu_3\mu _1;  \quad  I_3=\mu_1\mu _2\mu _3
\en
    The quadratic Casimir is zero for the pure state, the cubic zero for pure state also and on the boundary line QP. We have:
\begin{tabular}{|c|c|c|}
  \hline % after \\: \hline or \cline{col1-col2} \cline{col3-col4} ...
   & $I_2$ & $I_3$ \\
  \hline
  $O$ & 1/3 & 1/27 \\
  P & 1/4 & 0 \\
  Q & 0 & 0 \\
  \hline
\end{tabular}

\par

        There is an inequality between $I_2$ and $I_3$, because the cubic equation has to have real roots;
        see \cite{4}.

    One can also parameterize the states with two angles \cite{2}, with e.g.
\be x=  \sin^2\theta/2 \cos^2\phi/2, \q  y= \sin^2\theta/2
\sin^2\phi/2, \q z = \cos^2\theta/2 \en

\begin{figure}
  \includegraphics[width=9 cm]{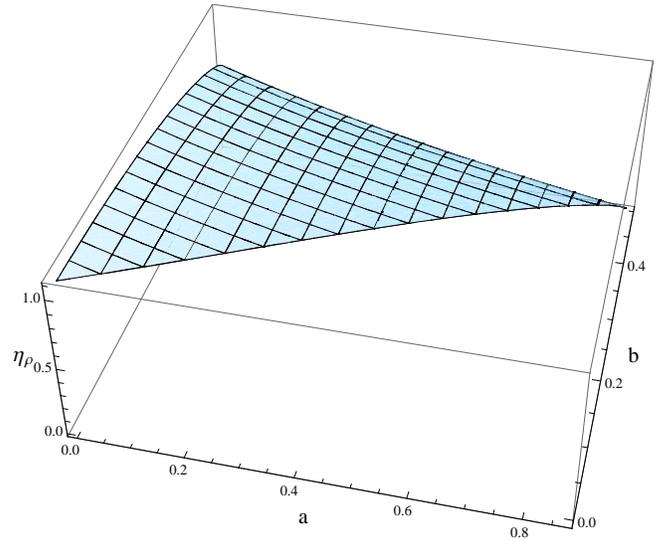}\\
  \caption{"(color online)" The entropy surface over the triangle $OPQ$}\label{fig:3dentropy}
\end{figure}

\begin{figure}
  \includegraphics[width=9 cm]{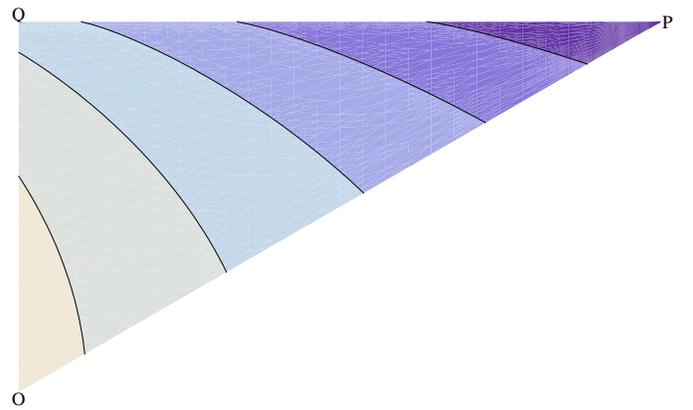}\\
  \caption{"(color online)" Constant entropy lines over the triangle $OPQ$}\label{fig:3disentropic}
\end{figure}

    The entropy of the ($xyz$) state, with $x+y+z=1$, is $\log(x^{-x} y^{-y}
    z^{-z})$. Fig \ref{fig:3dentropy} shows the entropy surface.
    It varies from log(3) at $O$ to log(2) at $Q$ to
    log(1)=0 at $P$.
    Fig \ref{fig:3disentropic} shows the isentropic lines over the
    triangle. The point $R$ on line $OP$ with
    $\eta(R)=\log(2)$ is $xyy=$.768, .116,  .116.

\begin{figure}
  \includegraphics[width=10 cm]{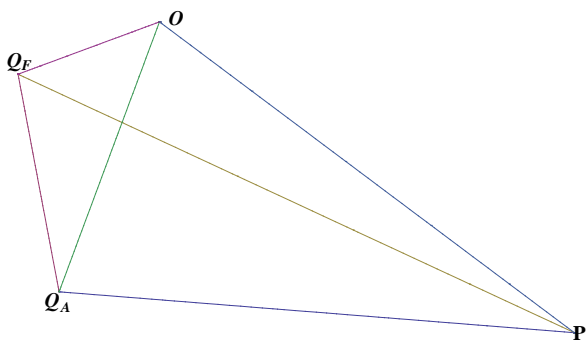}\\
  \caption{"(color online)" The little tetrahedron for $N$=4}\label{fig:th}
\end{figure}

% It is nice to plot the entropy surface, lying over the small triangle.It is nice also to draw the lines of constant entropy,
%h(line s) =cons. To compute: the point over the line OP with
% entropy log 2, as the Q state.
We have come now to the $\bf{N=4}$ case. The general state with
$\lambda_1, \ldots \lambda_{15}$ is
 \be
  \rho = \fr{1}{4} ( 1 + \sqrt{6} \: \bf{n} \cdot \vec{\lambda}).
  \en
We take $\lambda_{15}$ as
\begin{equation}
 \lambda_{15}= \frac{1}{\sqrt{6}} \mbox{diag} \{1,1,1,-3 \},
\end{equation}
 which gives the diagonal density matrix
 \begin{widetext}
\begin{equation}
\rho \approx \frac{1}{4}\left(\begin{array}{cccc} 1 +
\sqrt{6}a+\sqrt{2}b+c&  & 0 & 0 \\0 & 1-\sqrt{6}a+\sqrt{2}b+c & 0
& 0 \\0 & 0 & 1-2\sqrt{2}b+c & 0 \\0 & 0 & 0 & 1-3c \\
\end{array}\right)
\end{equation}
\end{widetext} in particular $c \le 1/3$. The figure is now a regular
tetrahedron inscribed in $S^2$. Again, the action of the 24
elements of the $S_4$ group generates the little rectangular
tetrahedron as quotient (fig \ref{fig:th}).

In the eigenvalue notation $(xyzu), \; x+y+z+u=1$, the coordinates
of the vertices are:

$O$ is (1/4, 1/4, 1/4, 1/4); $mM$ state=$U(4)/U(4)$, partition
[4].

$P$ is (1, 0, 0, 0); pure state=$U(4)/[U(1) \times U(3)]$,
partition [3,1].

$Q_A$ is (1/2, 1/2, 0, 0); nongeneric mixed state=$U(4)/U(2)^2$,
partition [2,2].

$Q_F$  is (1/3, 1/3, 1/3, 0); nongeneric mixed
state=$U(4)/[U(1)\times U(3)]$, partition [3,1].

There are six lines joining vertices, as follows
\begin{tabular}{|c|c|c|c|c|c|c|} \hline
 &$OP$&$OQ_A$&$OQ_F$&$Q_AP$&$Q_AQ_F$&$Q_FP$\\ \hline
  Length & 1 & $1/\sqrt{3}$ & 1/3 & $\sqrt{2/3}$ & $\sqrt{2}/3$ & $2 \sqrt{2}/3$ \\
  Spectrum & $x>yyy$ & $xx>yy$ & $xxx>y$ & $xy00$ & $xx>y0$ & $yy>x0$ \\
  type & [3,1] & [2,2] & [3,1] & [2,1,1] & [2,1,1] & [2,1,1] \\
  \hline
\end{tabular}

There are four faces, we describe just two here:
\begin{tabular}{|c|c|c|}
  \hline
  &$OPQ_A$ &  $PQ_FQ_A$ \\
  \hline
  Spectrum & $xyzz$ & $xyz0$ \\
  Type & [2,1,1] & [1,1,1,1] \\
  \hline
\end{tabular}.

Finally, there are five types of orbits:
\begin{widetext}
\begin{tabular}{|c|c|c|c|c|c|}
  \hline
  Spectrum & 1/4, 1/4, 1/4, 1/4 & $xxx, \; 1-3x$ & $xxyy$ & $xyzz$ & $xyzu$ \\
  \hline
  Representative & $mM$=O & $V$=1000 & $Q_A$ & $Q_AP$ & Interior \\
  dim & 0 & 6 & 8 & 10 & 12 \\
  character & Fixed point & Pure states $\approx CP^2$ &  &  & Generic (Flag manifold) \\
  G/H & U(4)/U(4) & U(4)/[U(1)xU(3)] & U(4)/U(2)$^2$ & U(4)/[U(2)xU(1)$^2$] & U(4)/U(1)$^4$ \\
  \hline
\end{tabular}
\end{widetext}
The discussion of the Casimir invariants and the calculation of
the entropy follows similar lines to the $N= 3$ case, so we omit
the details here. The face $Q_FQ_AP$ reproduces the previous $N=3$
case. This $N=4$ case is important as being the first case with
correlation, for example entanglement.

\section{GENERALIZATION}
The density matrix for an $N$ level system is given by \be
             \rho =   \frac {1}{N}  \left( 1 + \sqrt{\frac{N(N-1)}{2}} \vec{n }\cdot \vec{\lambda} \right)
\en where the $\lambda$'s are $N^2-1$ Hermitian traceless $N
\times N$ matrices with square 2; $N-1$ of them can be
diagonalized simultaneously; let us call them $\lambda_3,
\lambda_8, \lambda_{15}, ..., \lambda_{N^2-1}$. Then the density
matrix is given by
\begin{widetext}
\be
\rho=\fr{1}{N}\left(%
\begin{array}{cccc}
  1+\sqrt{\frac{N(N-1)}{2}} \left(a +b/\sqrt{3} + c/\sqrt{6} + \ldots + z \sqrt{\frac{2}{N(N-1)}} \right) & 0 & 0 & 0 \\
  0 &1 +\sqrt{\frac{N(N-1)}{2}} \left(-a + b/\sqrt{3} + \ldots \right) & 0 & 0 \\
  0 & 0 & \ldots & \ldots \\
  0 & 0 & \ldots & 1-(N-1)z \\
\end{array}%
\right) \en \end{widetext} So again, $-1 \leq z \leq 1/(N-1)$; the
ranges of eigenvalues always give us a regular simplex or solid
hypertetrahedron $T_N$, because there are $N+1$ orthogonal pure
states in $CP^N$, symmetrically distributed in the $S^{N-1}$
sphere of unit radius. The hyperfaces of the polytope delimit the
range. The center corresponds to the $mM$ state, which remains
invariant under $U(N+1)$ or $PU(N+1)$.

    The (full) symmetry group is $S_{N+1}$, of course (arbitrary permutation of vertices etc.); this divides $T_N$ in $N!$  rectangular, irregular little hypertetrahedra. In any of them we have $O$, the $mM$ state; $P$, the vertex; and $N-1$ $Q$'s, from $Q_F$ (center of the hyperface) to $Q_A$ (center of the edge uniting, in $T_N$, $P$ with another vertex $V'$).

    Notice in this picture some elements are interior in the original, regular polytope, others are at the boundary (extremal); for example, the cell which does not contain the $O$ = $mM$ is in the boundary, but the others are interior; similar for other elements: faces, edges, etc.

    The calculation of the Casimir invariants starting from eq. (23) is mechanical and trivial. Some general results are
\be
        I_N ( = det \rho) = 0 \mbox{ for a boundary state}
\en
because of a zero eigenvalue. Also
\be
            I_{N-1} = 0 \mbox{ for a boundary edge state}
\en
because of two zero eigenvalues, etc. Besides
\be
            I_j(mM \mbox{ state}) = \left(%
\begin{array}{c}
  N \\
  j \\
\end{array}%
\right) \frac{1}{N^j}
\en
because all the eigenvalues are equal to $1/N$.

    Let us discuss the entropy in this general situation; the formula
\be
        \eta(\rho) =\log( \Pi  \mu_i^{-\mu_i}),  \quad     \sum \mu_i = 1
\en
can be applied without any difficulty; again, we just include some results
\be
    \eta(O) = \log N > \eta(\rho) > \log (P) = \log 1 = 0
\en
for   $O  \neq \rho  \neq  P$.   Besides for $A$= edge, $F$= face, we have
\be
    \eta(Q_A) = \log 2, \;  \eta(Q_F) = \log 3, \; \ldots, \;  \eta(Q_{cell}) = \log (N-1)
\en where $Q_A$ is in the boundary edge, $Q_F$ in the boundary
face, ..., $Q_{cell}$ in the boundary cell, the $(N-1)$-polytope.
There are plenty of isentropic surfaces, which we refrain to make
explicit, as we already established them in the $N=3$ case. See
also \cite{2}.

    \section{FINAL  REMARKS}

    Our purpose in this work has been to describe the eigenvalue set and the orbit space of density matrices in a finite quantum system. The picture
    is
    \begin{eqnarray}
                               \{\mbox{set of states}\}   \leftrightarrow   \mbox{solid hypertetrahedron }T_N \\
 \{ \mbox{orbit space} \} = T_N/S_{N+1}  \leftrightarrow \no  \mbox{ rectangular} \\ \mbox{nonregular small \textit{h}-tetrahedron }t_N
\end{eqnarray}
where $t_N$ = $\{$set of density states$\}$/$PU(N)$.

    The actual geometry of mixed states, out of the $N=1, 2$ cases, is rather involved \cite{8} and we have not pretended to improve on it.
    There is also the interesting problem of finding maps between density states, positive/strictly positive, untouched also here;
    see e.g. the recent work \cite{9}.

    The geometry of the complex flag manifolds $U(N)/U(1)^N$ is very rich; we remark here only one result
\be
        U(3)/U(1)^3 := Fl_C (3) = SU(3)/U(1)^2  \sim CP^1  \odot CP^2
\en where the twisted product $\odot$ is similar to the used in
\cite{10}  to express the homology of Lie groups.

The $N=3$ case reminds one of the 3-dim Jordan algebras: the
action of $SU(3)$ on the 3 x 3 hermitian traceless matrices is
similar to the action of the exceptional group $F_4$ on the
exceptional Jordan octonionic algebra; indeed, the Moufang plane
$OP^2$ is $F_4/Spin(9)$ (Borel, 1950), whereas our $CP^2$ is
$SU(3)/U(2)$. In fact, the two cases are related, as the following
beautiful chain of groups show

\be
\begin{array}{ccccccc}
     &  & SU(3) &  & Spin(8) & & E_6 \\
     &&&&&&\\
   &  a\swarrow & d\uparrow & a\swarrow & t\uparrow & a\swarrow & \\
   &&&&&& \\
   SO(3)&  & G_2 &  & F_4 &  &  \\
\end{array}
\en where $a\equiv$Aut leads to the fixes sets (i.e.  $E_6
\rightarrow F_4$, $Spin(8)$  $\rightarrow G_2$ and $SU(3)
\rightarrow SO(3)$ ), and $t$ (trialitiy) and $d$ (duality) means
to relate Spin(8) with $F_4$ and $SU(3)$ with $G_2$. That is, $F_4
\approx Spin(8)$ plus the 3 8-dim representations and $G_2 \approx
SU(3)$ plus \textbf{3} and $\bf{\bar{3}}$.

\begin{acknowledgments}
We wish to thank Prof. Sudarshan, Prof. Byrd and Dr. Modi for
previous discussions. L. J. B. thanks M. E. C. (Spain) for grant
FPA-2006-20315.
\end{acknowledgments}

\end{document}